\RequirePackage[2020-02-02]{latexrelease}
\documentclass[showpacs,show keys,11pt,
preprint,preprint numbers,nofootinbib,
groupedaddress,superscriptaddress,amsmath,amssymb]{revtex4}
\usepackage{xcolor}
\usepackage{amsfonts}
\usepackage{amsmath}
\usepackage{graphicx,subfigure}
\usepackage{caption}
\usepackage[export]{adjustbox}
\usepackage{epsfig}
\usepackage{url}
\usepackage{multirow}
\usepackage{feynmp}

\newcommand {\be}{\begin{equation}}
\newcommand {\ee}{\end{equation}}
\newcommand {\ba}{\begin{eqnarray}}
\newcommand {\ea}{\end{eqnarray}}

\begin{document}
\title{Exploring Charged Higgs at the Future Circular Collider (FCC-hh): A Review of Two-Higgs-Doublet Models}

\author{Ijaz Ahmed}
\email{Ijaz.ahmed@fuuast.edu.pk}
\affiliation{Physics Department, Federal Urdu University of Arts, Science and Technology Islamabad, Pakistan}
\author{Basit Ali}
\email{basit4319@gmail.com}
\affiliation{Physics Department, Federal Urdu University of Arts, Science and Technology Islamabad, Pakistan}
\author{M.S. Amjad}
\email{sohailamjad@nutech.edu.pk}
\affiliation{National University of Technology, Islamabad}
\author{M. Jamil }
\email{mjamil@konkuk.ac.kr}
\affiliation{Department of Physics, Konkuk University, Seoul South Korea}
\author{Saba Shafaq}
\email{saba.shafaq@iiu.edu.pk}
\affiliation{International Islamic University Islamabad Pakistan}
\author{Usman Ahmad}
\email{usmanahmed1661@gmail.com}
\affiliation{Riphah International University, Islamabad Pakistan}
\date{\today}
\begin{abstract}

This paper presents a theoretical study of the charged Higgs bosons and their interactions with fermions in the Two Higgs Doublet Model (THDM).
We studied the decay, coupling to fermions, branching ratios, and production cross-section of charged Higgs bosons in the context of pp$\rightarrow$$tH^-$ interactions. Our analysis focused on identifying signatures of charged Higgs bosons and examining their kinematic distributions in the presence of background processes involving $b\bar{b}$ quarks. By considering two reference points and experimental limitations, we established an upper limit on the mass of charged Higgs bosons within the range of 200–1000 GeV.
%
We found that $H^- \rightarrow b\bar{t}$ and $H^- \rightarrow W^-A$ become high priorities in probing additional charged Higgs channels. We also presented the application of modern machine learning techniques such as Boosted Decision Tree (BDT), likelihood (LH), and Multilayer Perceptron (MLP) as an alternate approach to light-charged Higgs boson production in association with W boson in THDM Type-I via weak interaction, demonstrating its observability against the most relevant electroweak Standard Model backgrounds.
\end{abstract}
\maketitle
\section{Introduction}
The standard model (SM) of particle physics has provided a remarkably successful electroweak theory consistent with the current experimental observations. This was particularly recognized by the milestone discovery of the final constituent of the SM, the neutral spin-zero Higgs boson at the LHC in 2012 with a mass of about 125 GeV \cite{z1,z2}. Besides solving many longstanding fundamental problems of modern physics in light of
the predictions of the SM, such as unitarity of high energy scattering amplitude involving massive weak gauge bosons ($W^{\pm}$ and Z) and renormalization of the electroweak theory \cite{z3}, the Brout-Englert-Higgs (BEH) process was finally validated by the discovery of the Higgs boson scalar in \cite{z4}.\\
In the SM, the BEH mechanism not only explains the origin of mass for the gauge bosons ($W^{\pm}$ and Z) and charged fermions (quarks and leptons) via spontaneous electroweak symmetry breaking (EWSB), but it also ensures unitarity of high energy amplitude. The masses of the SM particles emerge from their coupling to the Higgs boson. This further established the fact that mass is not an intrinsic property of particles. However, particles (like photons) have zero mass because they don't interact with the Higgs field. 
The Higgs boson is an excitation of the Higgs field having a non-zero vacuum expectation value (vev). After the discovery of the Higgs boson, measuring its coupling to the SM particles via EWSB is the next priority.
Experimental sensitivity to the Higgs coupling is determined by measuring the Higgs production cross sections, decay modes, branching fractions, and total decay width. To validate the Higgs coupling to massive particles through Electroweak Symmetry Breaking (EWSB), it is essential to explore the Higgs sector beyond the Standard Model (SM) to search for new particles and Higgs bosons that interact with both the SM Higgs boson and beyond the Standard Model (BSM) particles.

%
Further, despite the marvelous success of the SM, some theoretical as well as experimental evidence suggest that the SM is only a low-scale effective theory of a more fundamental one at high energy, hence such evidence calls for the extension of the SM too.\\
The extension is also necessary to solve some existing puzzles in the SM such as an abundance of dark matter that is about 80 $\%$ of the total matter contents of the Universe, dark energy that renders accelerating expansion of the Universe, matter-antimatter asymmetry rendering abundance of baryonic matter, the hierarchy problem that deals with the huge mismatch between EW scale and the Planck scale, neutrino mass problem via neutrino oscillations, and quantization of gravity, etc. A discovery of a single or several extra Higgs bosons will mark the evidence of the extended Higgs sector and departure from the SM. 
In theories featuring two Higgs doublets, the presence of charged Higgs bosons $H^{\pm}$ is essential. Charged scalars are often anticipated in models that incorporate an additional scalar doublet in the SM Higgs sector with SU(2)$_L$. Hence, if they are discovered, it will be clear evidence for BSM physics.

%
\section{Two Higgs Doublet Model and Benchmarking}
The Two Higgs Doublet Model (2HDM) is a very simple approach to test the signatures of the extended Higgs sector. There are four types of 2HDMs.
In Type-I, all quarks and leptons interact with the $\Phi_2$ scalar field. In Type-II, down quarks and charged leptons get their masses from the $\Phi_1$ scalar field, while up quarks acquire mass through the $\Phi_2$ coupling. In Type-X (or Type-III), all charged leptons interact with $\Phi_1$, and all quarks interact with $\Phi_2$. While Type III is also known as the Lepton-specific model. In Type-Y (or Type-IV), down quarks get their masses from $\Phi_1$, while up quarks and charged leptons get their masses from $\Phi_2$. The 2HDMs models predict a total of five Higgs bosons: two CP-even (h and H), one CP-odd (A), and two charged Higgs bosons $H^{\pm}$.

%
%
In 2HDMs, there are two scalar
doublets having scalar fields $\Phi_1$ $\&$ $\Phi_2$ and
hyper-charge $Y = \frac{1}{2}$. For the detailed theory and
phenomenology of 2HDMs models, the interested readers are referred
to \cite{z5}.
As of now, 2HDM of Type-I and Type-II have attracted greater attention
\cite{z4, z5, z6, z7, z8, z9}, in particular with regard to the
searches of extended Higgs sector at the future circular
hadron-hadron collider (FCC-hh). The 2HDM Type-III has been also invoked by some researchers \cite{z10,z11,z12,z13} to investigate the signatures of Higgs bosons beyond the SM.
However, there is a lack of studies on the discovery potential of charged Higgs bosons in 2HDM Type-III and Type-IV via the channel pp$\rightarrow$$tH^-$ at the future circular hadron-hadron collider (FCC-hh) with a center of mass energy of $\sqrt{s}$ = 100 TeV.

The Future Circular Collider (FCC) project explores three collision scenarios: FCC-hh for hadron-hadron collisions (proton-proton and heavy ion), FCC-ee for electron-positron collisions, and FCC-eh for electron-hadron collisions \cite{fccweb,abada}. In the FCC-hh, each beam would have a total center-of-mass collision energy of 100 TeV, compared to the 14 TeV at the LHC, such a scenario was utilized in our work \cite{fccbeam}.

 This study focuses on investigating the signatures of charged Higgs bosons, including their decays, fermion couplings, branching ratios, and production cross-sections, in the presence of background processes involving $b \bar{b}$ quarks. 
We established an upper limit on the mass of the charged Higgs in the 200-1000 GeV range by utilizing two benchmark points and experimental constraints. This study focused on the same channel as in \cite{z5}, examining Type-I and Type-II 2HDMs with similar constraints.

We used two benchmark points and experimental constraints to set an upper limit on the mass of the charged Higgs in the 200-1000 GeV range. The same channel has been studied in \cite{z5} using Type-I and Type-II 2HDMs with comparable constraints.

%
%

The two benchmark points for the channels under investigation are $m_H^{\pm}$=$m_H$ $\& $ $M_A$=100 GeV as BP1 and $m_H^{\pm}$=$m_H$=$M_A$ as BP2. The preferred condition for experimentalists is the alignment limit $\sin(\alpha-\beta)\rightarrow 1$. For this limit, neutral scalar Higgs \emph{h} behaves like SM Higgs having a mass 125 GeV. For this condition, the heavy Higgs H in 2HDM is gauge phobic because couplings to \emph{W/Z} are suppressed. In this alignment limit the decay $H^{\pm}\rightarrow W^{\pm}\emph{h}$ vanishes completely. The LHC experiments at $\sqrt{s}$=8 TeV have established limits on the charged Higgs mass and $\tan\beta$ plane for $H^{\pm}\rightarrow\tau^{\pm}\nu$ \cite{54,55} and $H^{+}\rightarrow t\bar{b}$ \cite{56,57} decay modes. The range $m_H^{\pm}=80-160$ GeV is particularly constrained by the $H^{\pm}\rightarrow t\bar{b}$ decay channel.

%

The scheme to probe charged Higgs $H^{\pm}$ has mass dependence
that tells us about the total cross-section and also decay
available channels, which can be grouped in two types;
low and high mass cross-section scenarios. For low mass of charged
Higgs, if it satisfies the condition $m_t$ $>$$m_H^{\pm}$ then the
decay of charged Higgs $H^{\pm}$ to $\tau$$\nu$ channel becomes
dominant in the region where $\tan$$\beta$$>$1 and $H^+$ can be
produced via the t$\rightarrow$$H^+$$\bar{b}$ decay channel.
However for $\tan\beta>1$ decay \emph{t}$\rightarrow W^+ b$
becomes dominant for both 2HDM-I and 2HDM-II. Here we
have imposed the condition that $m_{H^\pm}$=$M_{W^\pm}$.
For the production cross-section, we have considered the process
pp$\rightarrow$\emph{t}$H^+$ calculated at FCC-hh Collider.
We have found that for large masses the decay mode
$H^+$$\rightarrow$t$\bar{b}$ is dominant. Our results
show that pp$\rightarrow$\emph{$\bar{t}$}$H^+$ is the dominant
channel for charged Higgs production given by
g$\bar{b}$$\rightarrow$$H^+$$\bar{t}$.
From Figure \ref{fig:cst3}, we can see a comparison of cross sections of charged Higgs production at different particle colliders. The charged Higgs boson production cross section in the 2HDM type III model varies significantly with the value of tan$\beta$, with enhanced production at high values of tan$\beta$ due to the increased mixing between the two Higgs doublets. 
The production cross-section is influenced by the charged Higgs boson mass and the production channel. In Type IV 2HDMs, the behavior with tan$\beta$ is non-monotonic, showing a minimum at intermediate tan$\beta$ values due to top quark decays. The precise location of this minimum is determined by the model's parameters.

%
\begin{figure}[ht]
  \centering
  \includegraphics[width=10cm]{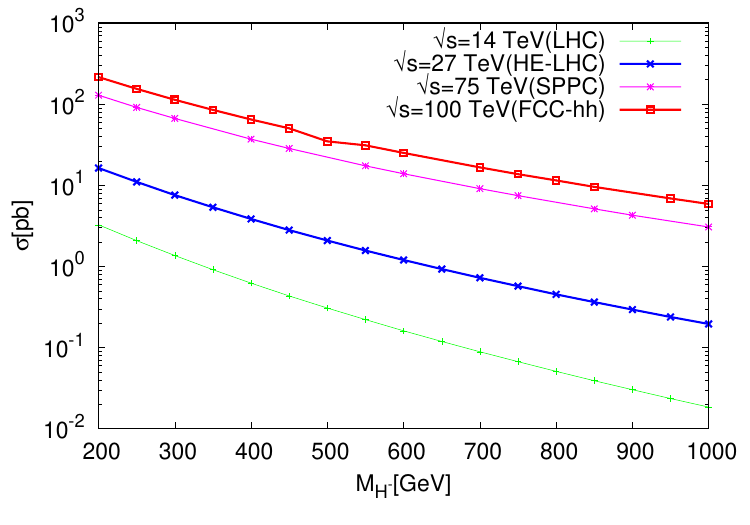}
\end{figure}
\section{Background Phenomenology}
The Lagrangian $\mathcal{L}$ of 2HDM is given as;
\begin{equation} 
\mathcal{L}_S=
\sum_{i=1}^2\left(\mathcal{D}_\mu\Phi_i\right)^\dagger\left(\mathcal{D}^\mu\Phi_i
\right)-V_{THDM}
\end{equation}
The covariant derivative is denoted as: 
\begin{equation}
\mathcal{D}_\mu=\partial_\mu-\frac{i}{2}g\sum_{\alpha=1}^{3}\sigma^\alpha\mathcal{W}_\mu^\alpha-\frac{i}{2}g'  \mathcal{B}_\mu
\end{equation}
The general Higgs potential is defined as;
\begin{equation}
\begin{split}
V_{THDM}=m_1^2|\phi_1^2|+m_2^2|\phi_2^2|-m_3^2|\phi_1^\dagger\phi_2|+h.c.\\
+\frac{1}{2}\lambda_1|\phi_1^4|+\frac{1}{2}\lambda_2|\phi_2^4|+\lambda_3|\phi_1^2||\phi_2^2|+\lambda_4|\phi_1^\dagger\phi_2|^2\\
+\frac{1}{2}\lambda_5(\phi_1^\dagger\phi_2)^2+\lambda_6|\phi_1|^2\phi_1^\dagger\phi_2+\lambda_7|\phi_2|^2\phi_1^\dagger\phi_2+h.c.
\end{split}
\end{equation}
The parameters $m_1$, $m_2$, $\lambda_1$-$\lambda_4$ are real, while $m_3$ and $\lambda_5$-$\lambda_7$ are complex. The
Higgs doublets are defined as;

\begin{equation}
\phi_1=\begin{pmatrix}
    \omega_1^+\\
    \frac{1}{\sqrt{2}}\left(\nu_1+h_1+iz_1\right)
\end{pmatrix}
\end{equation}
\begin{equation}
    \phi_2=\begin{pmatrix}
        \omega_2^+\\
        \frac{1}{\sqrt{2}}\left(\nu_2+h_2+iz_2\right)
    \end{pmatrix}
\end{equation}
In 2HDM, the charged Higgs boson can decay into a pair of fermions through Yukawa interactions. The partial decay width of this process depends specifically on the nature of the coupling between the charged Higgs boson and the fermions.
Type III has one doublet coupling to up-type quarks and leptons, while the other couples to down-type quarks. 
In the Type III Two-Higgs-Doublet Model, the charged Higgs boson $H^{+}$ couples to down-type quarks and charged leptons, but not to up-type quarks. To describe the decay of the charged Higgs boson into down-type quarks, we can use the appropriate coupling constants derived from the general framework of 2HDMs.
For the decay of a charged Higgs boson into down-type quarks, the relevant partial decay width can typically be written in terms of the mass of the charged Higgs $m_{H^{+}}$ and the mass of the down-type quark $m_{d}$ (or other down-type quarks). The decay width is influenced by the decay channels available and the structure of the Higgs couplings.
The average partial decay width for the process $H^{+} \rightarrow d_i \bar{u_j}$ (where $d_i$ represents the down-type quark and $\bar{u_j}$ is the antiquark of the up-type quark) can be approximated as:
\begin{equation}
\Gamma(H^{\pm} \to d_i \bar{u}_j) = \frac{3}{16\pi} \frac{g^2}{m_W^2} |V_{ij}|^2 \, p_{cm} \times \left[ m_{u_i}^2 \cot^2 \beta + m_{d_j}^2 \tan^2 \beta \right]
\end{equation}
where $p_{cm}=\frac{1}{2m_{H^{\pm}}}\sqrt{m^2_{H^{\pm}}}-(m_{u_i}+m_{d_j})^2 (m^2_{H^{\pm}}-(m_{u_i})-m_{d_j})^2), V_{ij}$ is the CKM matrix element corresponding to the transition between the down-type quark $d_i$ and the up-type quark $\bar{u_j}$ and tan$\beta$ is the ratio of vacuum expectation values of both Higgs doublet. 

The decay width for $H^+$ decaying into a charged lepton $l^+$ and a neutrino $\nu_l$ can be similarly expressed in Type III as:
\begin{equation}
\Gamma(H^+ \to l^+ \nu_l) = \frac{G_F}{4\sqrt{2}\pi} \frac{m_{H^+}^2}{v^2} \left( m_l^2 \right) \sqrt{1 - \frac{(m_l)^2}{m_{H^+}^2}}
\end{equation}
In the Type IV Two-Higgs-Doublet Model, the charged Higgs boson $H^+$ interacts primarily with down-type quarks and charged leptons. The relevant decay modes involve decays into a down-type quark and a down-type antiquark or a lepton and corresponding neutrino.
The partial decay width $\Gamma(H^+ \to f \bar{g}$) for the process $H^+ \to f \bar{g}$ (where f is a down-type quark or leptons and $\bar{g}$ is another anti-down-type quark or anti-neutrino, respectively) can be expressed generically in terms of the Yukawa coupling strength, mass, and phase space factors.
For example, for the decay $H^+ \to t \bar{b}$ (where t is the top quark and $\bar{b}$ is the anti-bottom quark) the partial decay width can be given by:
\begin{equation}
   \Gamma(H^+ \to t \bar{b}) = \frac{g^2}{32\pi m_{H^+}^2} \sqrt{\left(1 - \frac{(m_t + m_b)^2}{m_{H^+}^2}\right)\left(1 - \frac{(m_t - m_b)^2}{m_{H^+}^2}\right)} \times \left(m_t^2 + m_b^2\right) \times \left(1 + \delta_{g}\right)
\end{equation}
where, g is the relevant coupling constant, $m_{H^+}$ is the mass of the charged Higgs boson, $m_t$ and $m_b$ are the masses of the top quark and bottom quark, respectively, $\delta_{g}$ represents higher-order corrections or additional factors related to the couplings in the specific 2HDM scenario. Similarly, for $H^+$ decay into charged leptons, you would have:
\begin{equation}
  \Gamma(H^+ \to \ell^+ \nu_\ell) = \frac{g^2}{8\pi m_{H^+}^2} m_\ell^2 \sqrt{1 - \frac{m_\ell^2}{m_{H^+}^2}}  
\end{equation}
where $m_l$ is the mass of the charged lepton. The total decay width for the charged Higgs boson involves summing these contributions from all allowed decay channels. The precise values of the coupling constants and the mass terms would depend on the specific parameters of the Type IV 2HDM under consideration.

\section {Charged Higgs Couplings}
In 2HDM, there are two Higgs doublets, and the charged Higgs boson arises as a result of the mixing between the two doublets. The strength of the interaction between the charged Higgs boson and other particles is determined by the charged Higgs coupling constants.
In the Type III and Type IV variants of the 2HDM, the charged Higgs coupling constants are determined by the following formulas based on the masses of the particles.

\subsection{Type III}
The charged Higgs boson ($H^\pm$) interacts differently with quarks, leptons, and gauge bosons than the Standard Model (SM). The charged Higgs coupling relations in 2HDM type III can be expressed through a general Type-III charged Higgs-fermion interaction Lagrangian:

\begin{equation}
\label{eq:charged_higgs_quark_lagrangian}
\mathcal{L}_{H^{\pm} q_i \bar{q}_j} = \frac{g}{\sqrt{2}M_W} H^+ \left[ \bar{u}_i (V_{ij} \mathbf{X}_{ij}^L P_L + V_{ij} \mathbf{X}_{ij}^R P_R ) d_j \right] + \text{h.c.}
\end{equation}
where.
\begin{align}
\label{eq:10_revised}
\mathbf{X}_{ij}^L &= \frac{m_{u_i}}{M_W} \cot\beta \, \mathbf{\Delta}_{ij}^U \\
\label{eq:11_revised}
\mathbf{X}_{ij}^R &= \frac{m_{d_j}}{M_W} \tan\beta \, \mathbf{\Delta}_{ij}^D
\end{align}
where $\mathbf{\Delta}_{ij}^U$ and $\mathbf{\Delta}_{ij}^D$ are the dimensionless, complex matrix elements that encapsulate the specific flavor structure of the Type-III model.

\subsubsection{Leptons}
The charged Higgs boson couples to leptons through their Yukawa couplings with the first Higgs doublet, and the coupling strength is given by:
\begin{equation}
   g_{lH^\pm} = - \frac{m_l}{\sqrt{2} v_1}
\end{equation}
\subsubsection{Gauge bosons}
   The charged Higgs boson couples to gauge bosons through their mixing with the W boson, which depends on the mass of the charged Higgs boson and the mixing angle $\beta$. The coupling strength is given by:
\begin{equation}
   g_{W^\pm H^\pm} = - \frac{g}{\sqrt{2}} (1 + \tan^2\beta) + \frac{g m_{W^\pm}}{m_{H^\pm}^2} (1 + \tan^2\beta) \sin(2\beta)
\end{equation}
   where $g$ is the SU(2) gauge coupling constant.
\subsection{Type IV}
The charged Higgs boson interacts with up-type quarks ($u$ and $c$) and their corresponding charged leptons ($e$ and $\mu$) through Yukawa couplings in this model. The strength of these couplings is determined by:
\subsubsection{Up-type quarks}
Light quarks ($u$ and $c$):
\begin{equation}
    g_{H^\pm u\bar{d}} =  \frac{m_u}{v_1}\sin(\beta-\alpha) + \frac{m_c}{v_2}\cos(\beta-\alpha) 
\end{equation}
Heavy quarks ($t$):
\begin{equation}
    g_{H^\pm t\bar{b}} = -\frac{m_t}{v_2}\cos(\beta-\alpha) 
\end{equation}

In this model, $v_1$ and $v_2$ are the vacuum expectation values (VEVs) of the two Higgs doublets, $\beta$ is the mixing angle between the doublets, and $\alpha$ is another mixing angle. The first term in each coupling comes from the type I Yukawa sector, and the second term comes from the type II Yukawa sector.

\subsubsection{Charged leptons}
Light leptons ($e$ and $\mu$):
\begin{equation}
   g_{H^\pm e\bar{\nu}_e} = -g_{H^\pm \mu\bar{\nu}_\mu} = \frac{m_e}{v_1}\sin(\beta-\alpha) + \frac{m_\mu}{v_2}\cos(\beta-\alpha) 
\end{equation}
Heavy leptons ($\tau$):
\begin{equation}
    g_{H^\pm \tau\bar{\nu}_\tau} = -\frac{m_\tau}{v_2}\cos(\beta-\alpha) 
\end{equation}
\subsubsection{Gauge bosons}
   The charged Higgs boson also interacts with gauge bosons (W and Z) through gauge interactions. The coupling strengths are given by:
\begin{equation}
 g_{H^\pm W^\mp} = -\frac{g}{\cos\theta_W}(\sin(\beta-\alpha)-\cos(\beta-\alpha)) 
\end{equation}
\begin{equation}
   g_{H^\pm Z} = -\frac{g}{\cos\theta_W}(\sin(\beta-\alpha)\cos^2\theta_W+\cos(\beta-\alpha)\sin^2\theta_W)
\end{equation}
 The weak coupling constant $g$ and weak mixing angle $\theta_W$ depend on the mixing angles $\alpha$ and $\beta$.
 
\section{Results and Discussions}
 We used CalcHEP \cite{calchep1,calchep2} to calculate the decay width, branching fractions, and production cross-section for benchmark points BP-I and BP-II, including all interaction vertices. Gnuplot \cite{gnuplot} is a tool used to plot and visualize data and mathematical functions.
 
 In our calculations, we assumed the alignment limit $\sin(\beta-\alpha) = 1$, making the lighter Higgs with a mass of 125 GeV behave like the SM Higgs.\\
 The results include total decay width, branching ratios, and kinematic distributions of bottom (anti-bottom) quarks from background processes, presented in Figure \ref{fig:6fab} to Figure \ref{fig:mlp}.

\subsection{Total Decay Width of Charged Higgs}
\begin{figure}
\centering     
\subfigure[]{\label{fig:a}\includegraphics[width=75mm]{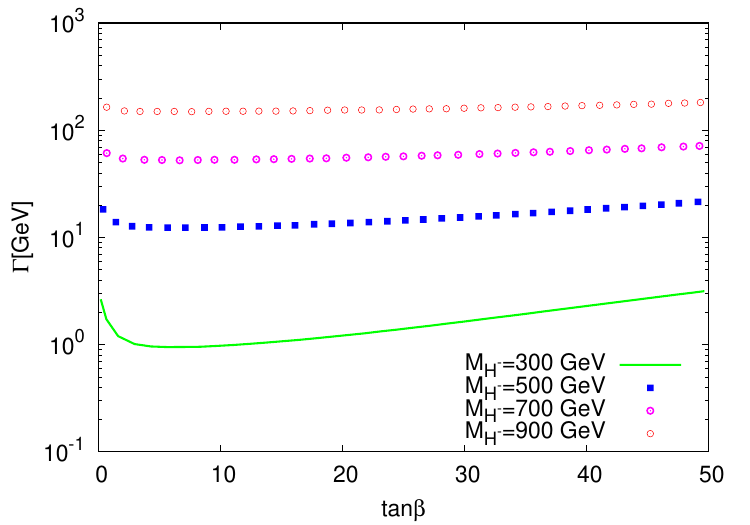}}
\subfigure[]{\label{fig:b}\includegraphics[width=75mm]{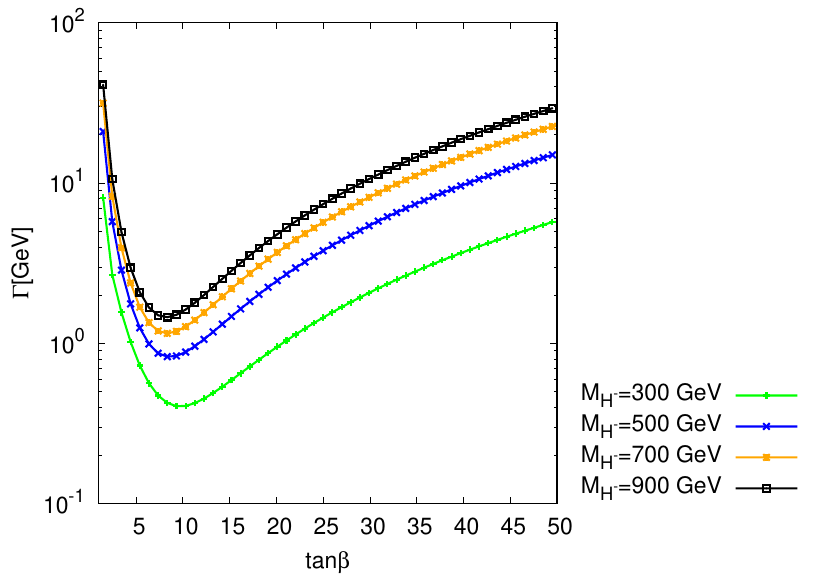}}
\caption{Plot is shown between $M_{H^\pm}$ and $\Gamma$ for $\tan\beta=1$.
Here, the Plot (a) shows the decay width of charged Higgs to all possible decay modes for BP-I. While Plot (b) shows BR of charged Higgs to all possible decay modes for BP-II}
\label{fig:6fab}
\end{figure}

The total decay width of the charged Higgs for parameters BP-I and BP-II is illustrated in Figure \ref{fig:6fab} and Figure \ref{fig:6f56} for THDM type-III and type-IV models. 
The figures indicate that as the mass of the charged Higgs increases beyond 300 GeV, the change in decay width is minimal. Additionally, for high $\tan\beta$ values, the decay width remains relatively constant.

For tan$\beta$=7, the decay width is minimum e.g using both BP-I and BP-II for THDM-II, and for the mass of charged Higgs $m_{H^-}$= 300 GeV, the value obtained for decay width is given as 0.35 GeV, which is illustrated in Figure \ref{fig:6fab}.
\subsection{Charged Higgs Branching Ratios of THDM-III}
\begin{figure}
\centering     
\subfigure[]{\label{fig:a}\includegraphics[width=81mm]{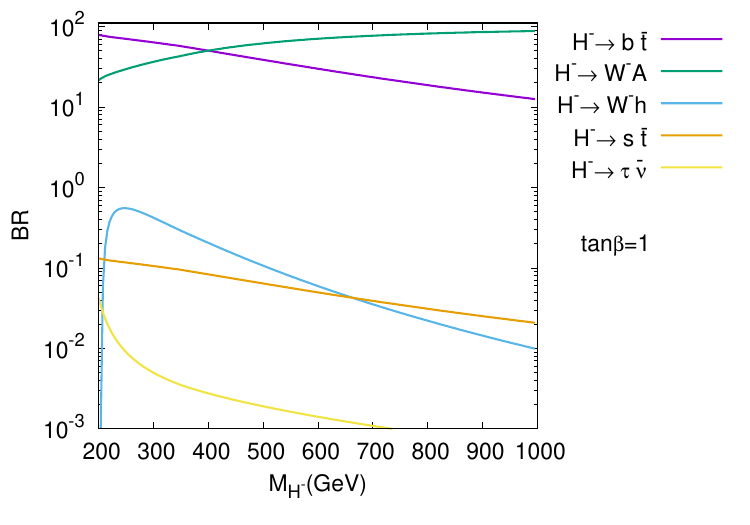}}
\subfigure[]{\label{fig:b}\includegraphics[width=81mm]{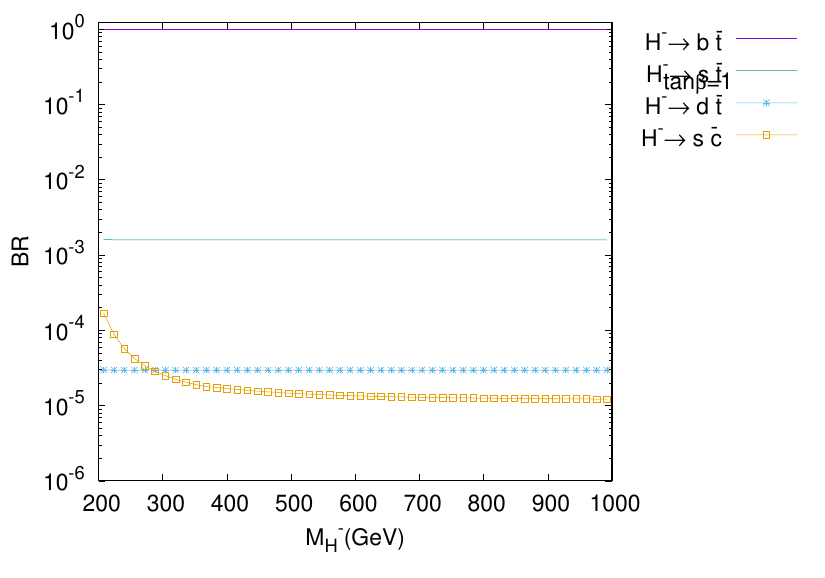}}
\caption{Plot between $M_{H^\pm}$ and BR in \% for $\tan\beta=1$.
Plot (a) shows the BR of charged Higgs to all possible decay modes for
BP-I. Plot (b) shows BR of charged Higgs to all possible decay modes for BP-II}
\label{fig:6f56}
\end{figure}

Figure \ref{fig:6f56} shows the relationship between the mass of the charged Higgs and the branching fractions for BP-I and BP-II, respectively.
To understand these decay channels of charged Higgs in more detail, graphical analysis is always there, by putting all data in Gnuplot \cite{gnuplot}, the graph is plotted between $M_{H^\pm}$ and BR which is shown in  Figure 3(b). It is obvious decay that $H^- \rightarrow b\bar{t}$ is highly dominant which is almost 100$\%$.

\subsection{Charged Higgs Branching Ratios of THDM-IV}
\begin{figure}
\centering     
\subfigure[]{\label{fig:a}\includegraphics[width=80mm]{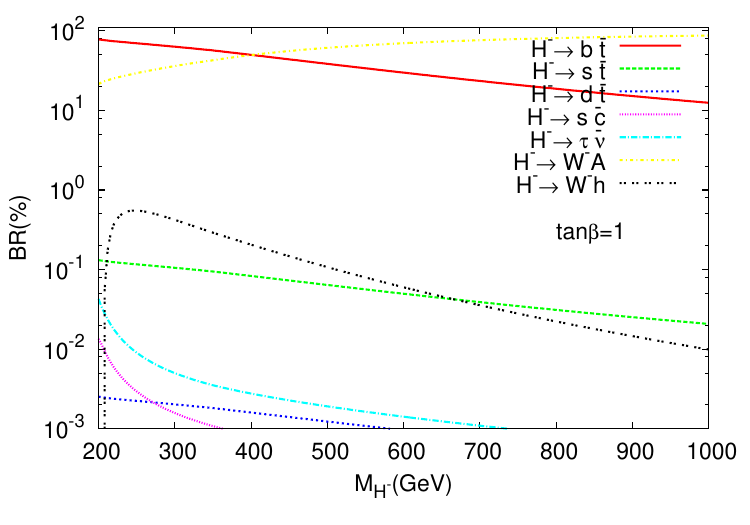}}
\subfigure[]{\label{fig:b}\includegraphics[width=80mm]{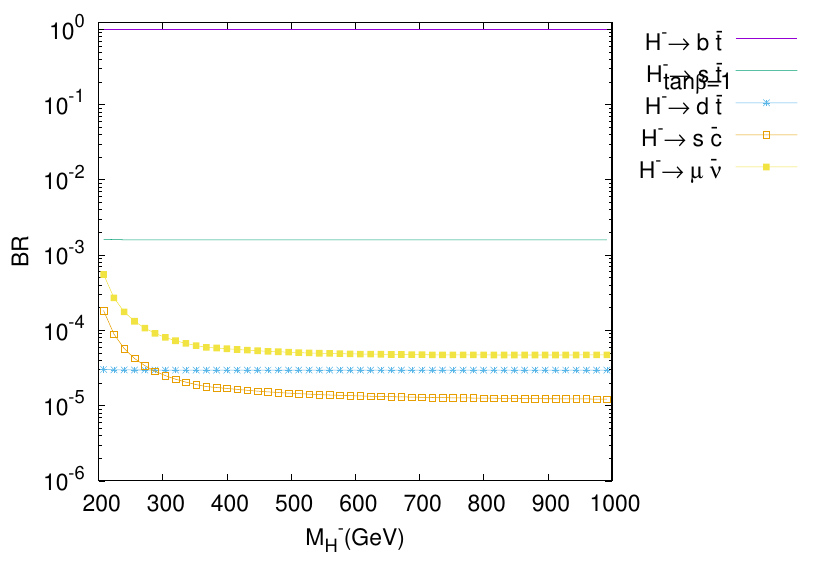}}
\caption{Plot between $M_{H^\pm}$ and $\%BR$ for $\tan\beta$=1.
Fig. (a) shows BR of charged Higgs for all possible decay modes for
BP-II. Fig. (b) shows BR of charged Higgs to all possible decay modes
for BP-I}
\label{fig:6f7}
\end{figure}

In THDM-IV, a similar situation occurs as in THDM-III, where the dominant decay mode is the $H^- \rightarrow
W^- A$ channel. This occurs when $M_{A^o} < M_t$, allowing for the opening of a bosonic channel.
Hence decay to this channel becomes
dominant. To illustrate this graphically, the values are plotted
between $M_{H^\pm}$ and BR which is shown in Figure \ref{fig:6f7}.
for BP-II (left) and BP-I (right).
Here we can see that dominant decay is again \emph{b}$\bar{t}$ channel and has BR up to 91$\%$. The BR to
other decay channels is almost constant, which shows the
independence of BR from $\tan$$\beta$. The b$\bar{t}$ is dominant
because top quarks have the largest mass of all fermions. Hence for
Higgs mass greater than top quarks masses the decay channel
\emph{b}$\bar{t}$ becomes dominant. All rest decay modes are
lowered or we can say that they are suppressed.
\section{Comparative Results}

\begin{figure}
\centering
    \includegraphics[width=0.5\linewidth]{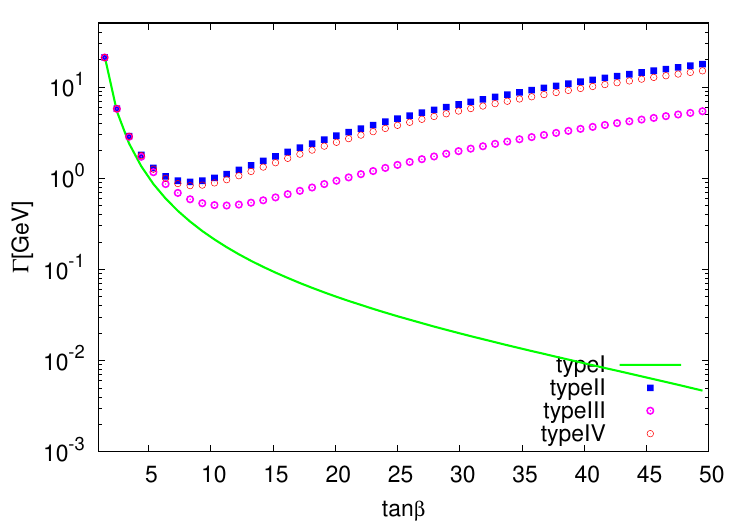}
    \caption{Total Decay width of charged Higgs as a function of $\tan\beta$ for $m_H^{\pm}$ = 500 GeV of all THDMs}
    \label{fig:mh500}
    \label{fig:5}
\end{figure}

\begin{figure}
\centering     
\subfigure[]{\label{fig:a}\includegraphics[width=81mm]{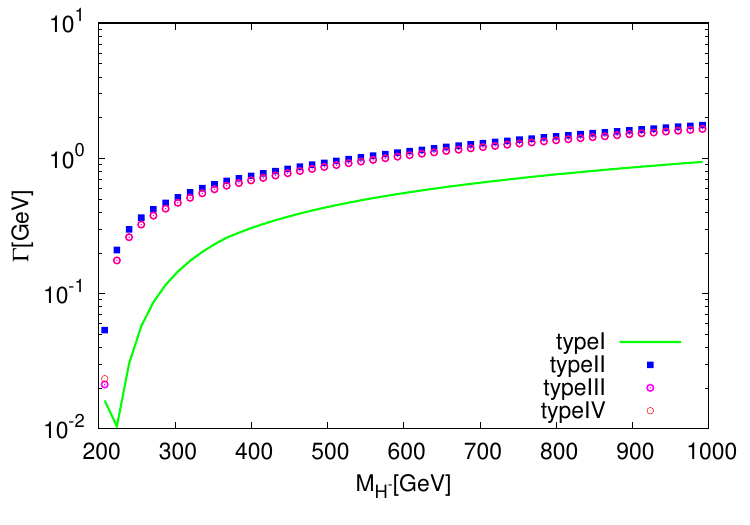}}
\subfigure[]{\label{fig:b}\includegraphics[width=81mm]{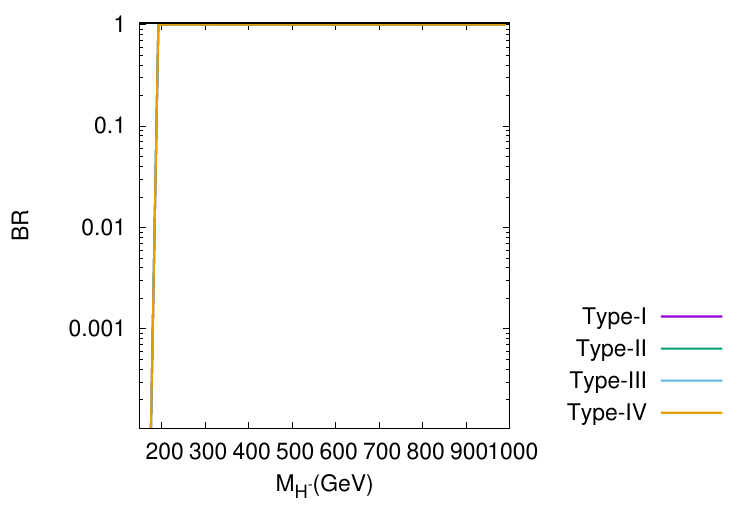}}
\caption{(a) The total Decay width of charged Higgs as a function of $m_H^{\pm}$ for $\tan\beta$ = 7 of all THDMs. (b) Plot between \%BR and $m_H^{\pm}$ for the decay $H^- \rightarrow b\bar{t}$ of all THDMs}
\label{fig:6}
\end{figure}

This section provides a comparison of the plots for all
THDMs. Figures \ref{fig:mh500} and \ref{fig:6} display the total decay width of charged Higgs as a function of $\tan\beta$ for $m_H^{\pm}$ = 500 GeV. It can be seen that the decay width for Type-III and Type-IV have the same values as expected from couplings. Because for both type-III and type-IV the quarks couples to the same type of scalar field. The decay width for type-IV has a little bit of less value this is because leptons couples to another type of scalar
field which is inverse of $\tan$$\beta$ i.e $\cot$$\beta$.\\

For type-I and type-III, such situation differs between type-I and type-III scenarios. In type-I, all fermions couple to the same scalar field, leading to a decrease in decay width as $\tan\beta$ increases. In contrast, type-III has a lower decay width compared to type-II and type-IV, but it increases due to leptons coupling with the same scalar field.


In Figure \ref{fig:6}, the total decay width of the charged Higgs is plotted against its mass for a tan$\beta$ value of 7 in all THDMs. 
The decay width for THDM type-I is minimal, as expected based on coupling relations. In type III, the decay width is slightly less than in type II and type IV. However, since leptons couple to the same type of field in type III, the values are almost the same as in type III and type IV.

In Figure \ref{fig:7}, a plot between \%BR and mass of charged Higgs for the decay $H^- \rightarrow b\bar{t}$ of all THDMs is presented. It can be seen clearly that it is the dominant decay mode of all THDMs, the graphic line overlap for all THDMs which is clear evidence that the dominant decay mode is b$\bar{t}$ channel of all THDMs.
\section{Production Cross sections}
In particle physics, the cross-section measures the likelihood of two particles interacting with each other during a process.
In a very simple way, the cross-section can be defined as an effective area in which the reaction will take place.
\begin{figure}
\centering     
\subfigure[]{\label{fig:b}\includegraphics[width=81mm]{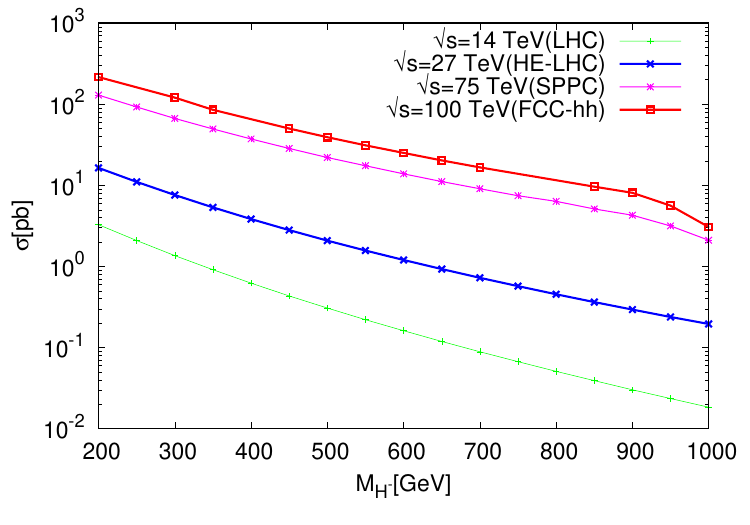}}
\caption{Plot between $\sigma$ and $M_{H^\pm}$ at different future colliders for THDM-III (left) and THDM-IV (right)}
\label{fig:7}
\end{figure}
From Figure \ref{fig:7}, we can also see the same situation, the cross-section decreases for large $\tan$$\beta$ values. However, its value is maximum for $\tan$$\beta=50$ and $60$. It can be seen from coupling relations that increasing $\tan$$\beta$ values increases the down quarks coupling values and leptons too. At large $\tan$$\beta$ leptons contribute to significant cross-section values, which results in greater cross-section values. Thus, the cross-section has units of area and in nuclear and particle physics the following definition is used, where 1 barn = $10^{-24} cm^2$.
A more detailed explanation can be found in \cite{67}.\\
Here in Figure \ref{fig:6} - Figure \ref{fig:7}, the plots between production cross section as a function of mass presented for the models THDM type-I, type-II, type-III, and type-IV for the process $pp\rightarrow tH^+$, for different colliders i.e LHC ($\sqrt{s}$ = 14 TeV), HE-LHC ($\sqrt{s}$ = 27 TeV), SPPC($\sqrt{s}$ = 75 TeV) and FCC-hh($\sqrt{s}$ = 100 TeV).\\

The cross-section plots show that the production cross-section is highest for FCC-hh due to its higher center of mass energy, offering a wide parameter space for charged Higgs detection. The same applies to SPPC, which also has a larger cross-section.

\section{Transverse Momentum Distributions}
It is necessary to measure the momentum of particles that are produced when particles collide. Now momentum has two components a transverse and longitudinal component.
\begin{figure}
\centering     
\subfigure[]{\label{fig:a}\includegraphics[width=81mm]{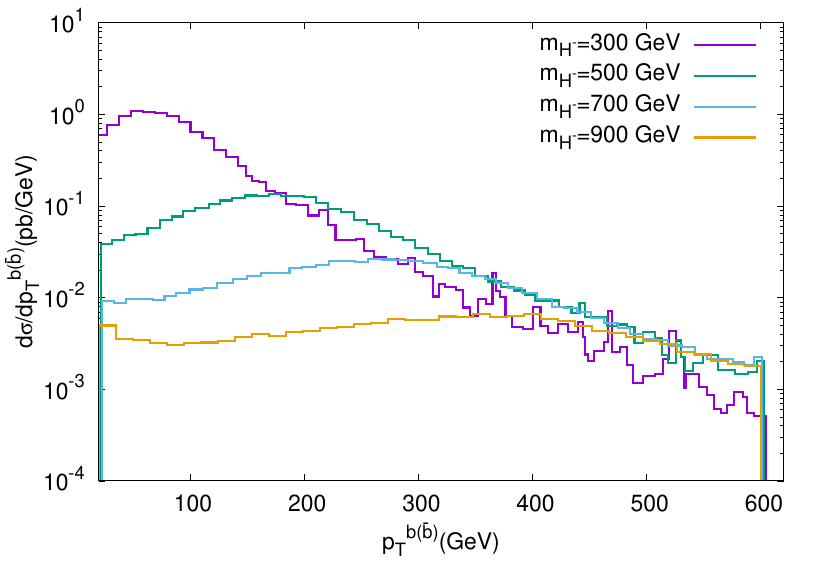}}
\subfigure[]{\label{fig:b}\includegraphics[width=81mm]{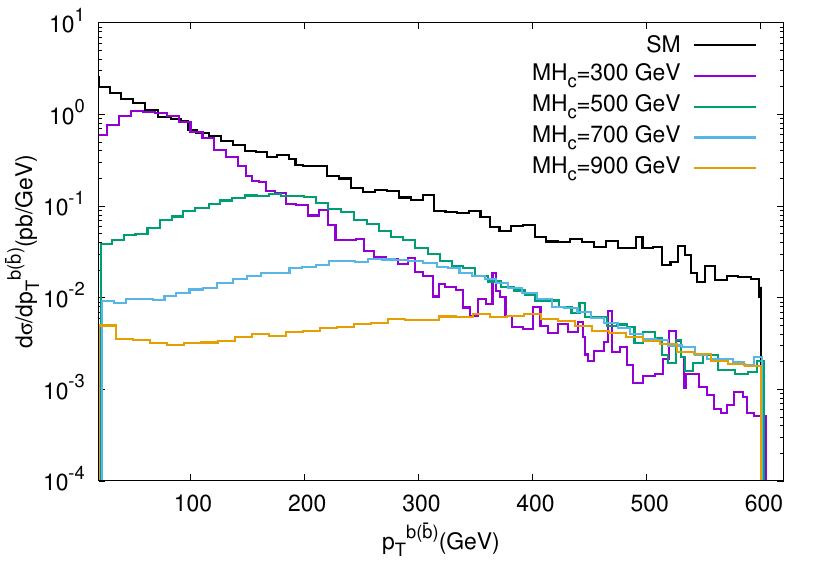}}
\caption{$P_T$ distributions of b($\bar{b}$) quarks  of THDM-III.
Fig. (a) shows transverse momentum distributions for BP-I and Fig. (b) shows transverse momentum distributions for BP-II }
\label{fig:8}
\end{figure}
The significant property of the transverse component is that it is invariant under Lorentz transformation. It is denoted by $p_T$. \\
The following conventions are adopted to obtain Cartesian momentum $p_x, p_y$, and $p_z$.
\begin{equation}
    p_x=p_T \cos\phi
\end{equation}
\begin{equation}
    p_y=p_T \sin\phi
\end{equation}
\begin{equation}
    p_z=p_T \sinh\eta
\end{equation}
\begin{figure}
\centering     
\subfigure[]{\label{fig:a}\includegraphics[width=81mm]{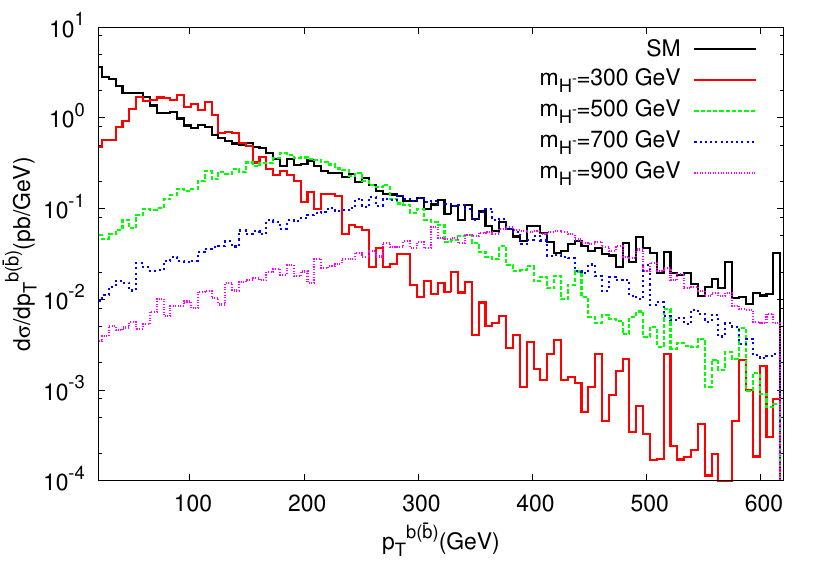}}
\subfigure[]{\label{fig:b}\includegraphics[width=81mm]{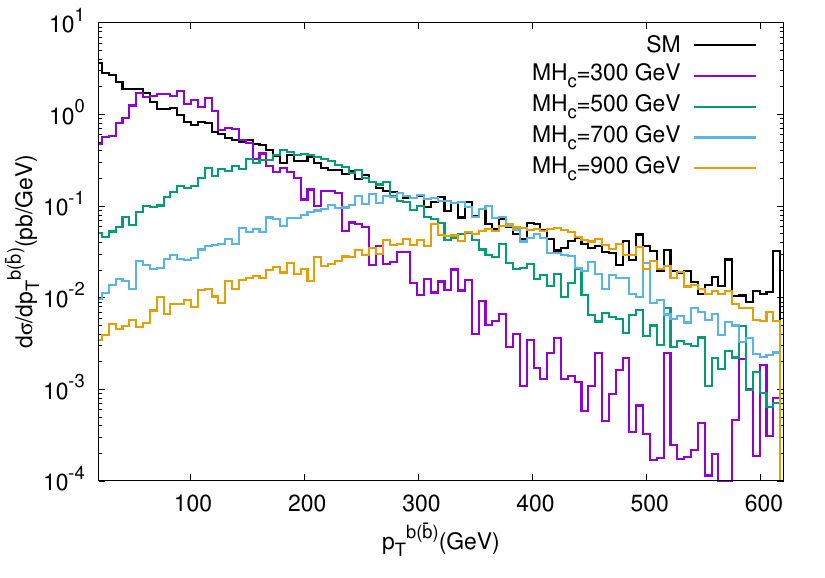}}
\caption{$P_T$ distributions of $b(\bar{b})$ quarks in THDM-IV are shown in Fig. 9. Figure 9(a) displays the transverse momentum distributions for BP-I, while Figure 9(b) shows the transverse momentum distributions for BP-II.}
\label{fig:9}
\end{figure}
To differentiate between signal and background events, three constraints are imposed on jets.
They are kinematics cuts on integrated jets, cuts on the b-jets, and cuts on aggregate di-jets. Now a lower limit is selected that four Jets must be detected in final states with cuts on $p_T and \eta$ due to uncertainties that come from the efficiency of b-tagging.\\
These cuts are given in equation (\textbf{21)}
\begin{equation}
    p_T^{Jets}\ge 20 \ GeV, |\eta^{Jets}|\le 3
\end{equation}
The $\rho_T$ distributions of b-jets for various charged Higgs masses are illustrated in Figure \ref{fig:8}(a-b) and Figure \ref{fig:9}(a-b).\\

\section{Pseudorapidity}
Pseudorapidity is an important kinematical variable, and it defines the angle of particle w.r.t beam axis, and it is given by
\begin{equation}
\eta=-\ln\left[\tan(\frac{\theta}{2})\right]
\end{equation}

The angle $\theta$ is defined as the angle between the direction of the beam axis and the momentum of the particle, and it is represented as:
\begin{equation}
\theta=2\tan^{-1}(e^{-\eta})
\end{equation}
In terms of the particle's three momenta, the $\eta$ can be expressed as
\begin{equation}
\eta=\frac{1}{2}\ln{\left(\frac{|p|+p_z}{|p|-p_z}\right)}
\end{equation}
From equation(24) we can write as;
\begin{equation}
e^\eta=\sqrt{\frac{|p|+p_z}{|p|-p_z}}
\end{equation}
$\&$
\begin{equation}
e^{-\eta}=\sqrt{\frac{|p|-p_z}{|p|+p_z}}
\end{equation}
From the addition of the above two equations we obtain
\begin{equation*}
|p|=p_T\cosh\eta
\end{equation*}
where $p_T$ is transverse momentum and is given as
\begin{equation*}
p_T=\sqrt{p^2-p_z^2}
\end{equation*}
\begin{figure}
\centering     
\subfigure[]{\label{fig:a}\includegraphics[width=81mm]{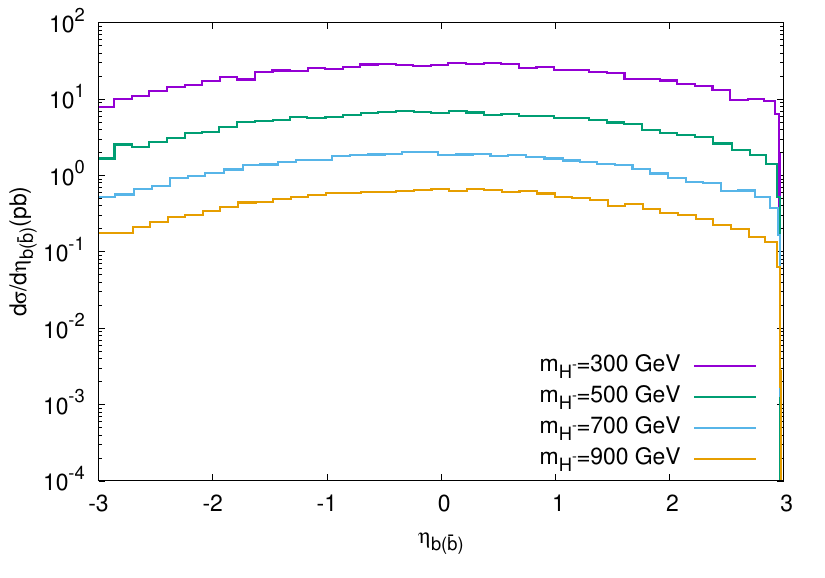}}
\subfigure[]{\label{fig:b}\includegraphics[width=81mm]{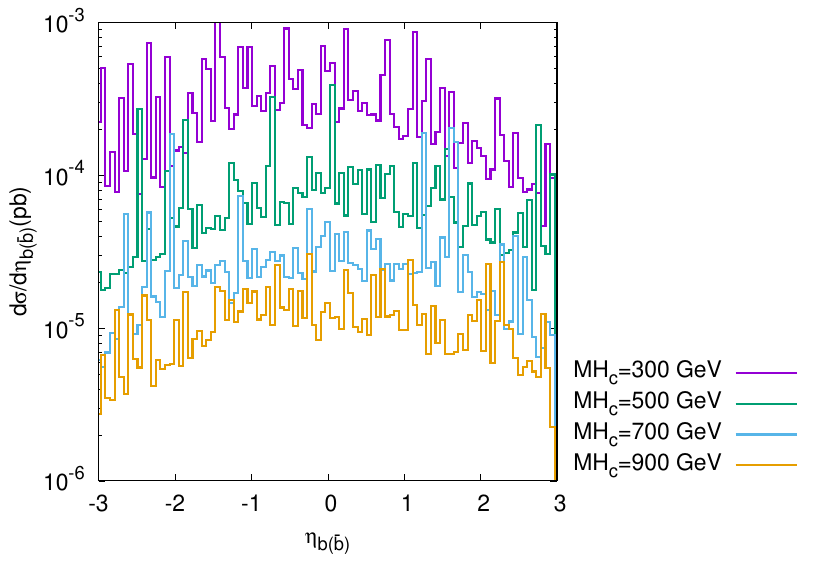}}
\caption{$\eta$ distributions of b($\bar{b}$) quarks of THDM-III.
Fig.a shows pseudorapidity distribution curves of b($\bar{b}$)
quarks for  BP-I. Fig. b shows pseudorapidity distributions of
b($\bar{b}$) quarks for  BP-II} 
\label{fig:10}
\end{figure}
More extensive detail can be found in text  \cite{66}.
In the center of the mass frame, the peak of the distribution is observed at \emph{y} $\approx$ $\eta$$ \approx$0.
\textbf{Figures \ref{fig:10}(a) and \ref{fig:10}(b)} show the \emph{b}($\bar{b}$) quark pseudorapidity distributions for THDM-I and THDM-II models.

\begin{figure}
\centering     
\subfigure[]{\label{fig:a}\includegraphics[width=81mm]{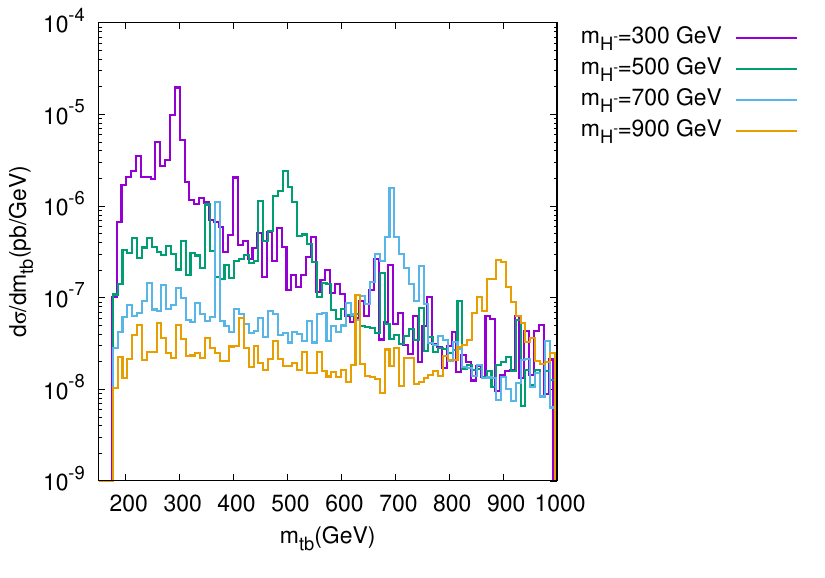}}
\subfigure[]{\label{fig:b}\includegraphics[width=81mm]{mtm4p1.pdf}}
\caption{$\eta$ distributions of b($\bar{b}$) quarks  of THDM-IV.
Fig a. shows pseudorapidity distribution curves of b($\bar{b}$)
quarks for  BP-I. Fig. b shows Pseudorapidity distributions of
b($\bar{b}$) quarks for BP-II}
\label{fig:11}
\end{figure}
In Figure 10(a) displays the distribution curves for BP-I of THDM-I, while Figure 10(b) shows the distributions for BP-II of THDM-I. Figure 10(b) illustrates the pseudorapidity distributions for BP-I, and in (b), the distributions are shown for BP-II of THDM-II. The distribution decreases when we go towards higher masses, which shows that massive particles are less deflected than lighter particles.
Also when peaks are obtained when $\eta$$\approx$0.
\section{Invariant Mass}
\begin{figure}
\centering     
\subfigure[]{\label{fig:a}\includegraphics[width=81mm]{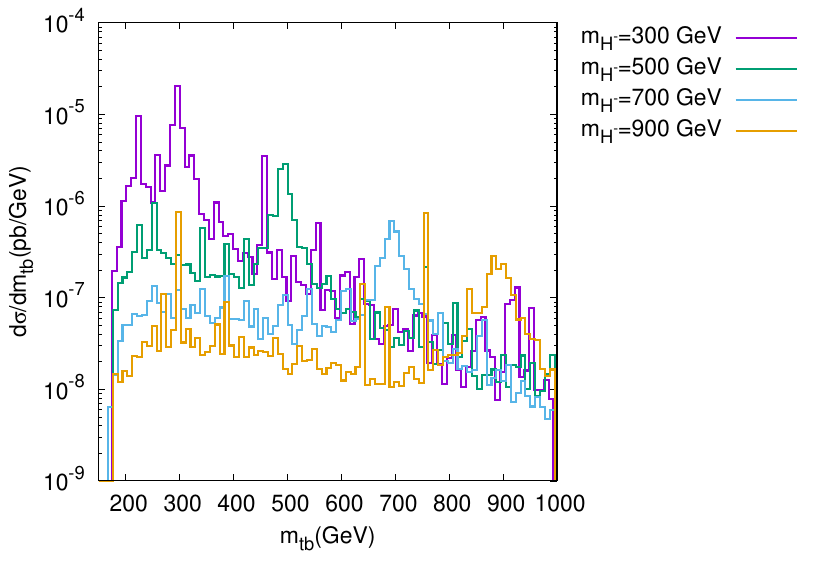}}
\subfigure[]{\label{fig:b}\includegraphics[width=81mm]{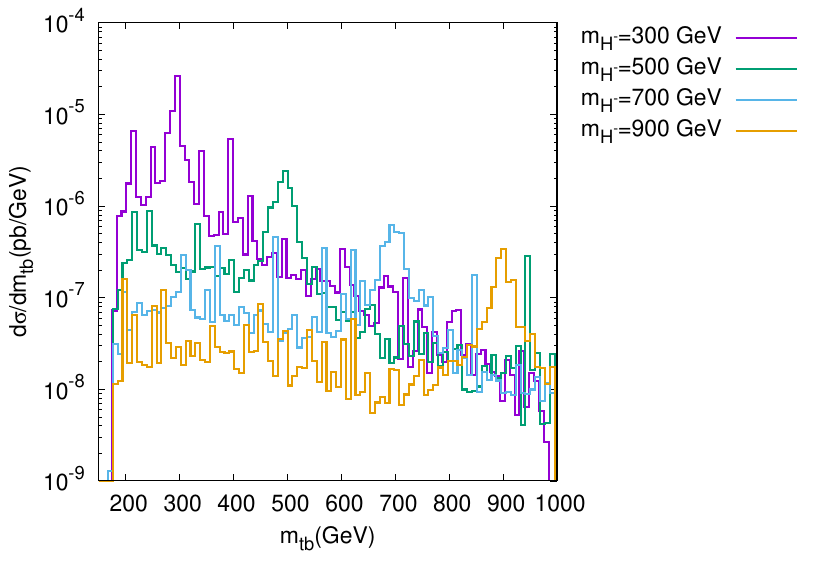}}
\caption{A view of the THDM-III invariant mass distributions of t$\bar{b}$ quarks. Fig.a shows $m_{tb}$ data plots of t$\bar{b}$ quarks for
BP-I. Fig.b shows $m_{tb}$ of t$\bar{b}$ quarks for BP-II}
\label{fig:12}
\end{figure}

\begin{figure}
\centering     
\subfigure[]{\label{fig:a}\includegraphics[width=81mm]{mtm4p1.pdf}}
\subfigure[]{\label{fig:b}\includegraphics[width=81mm]{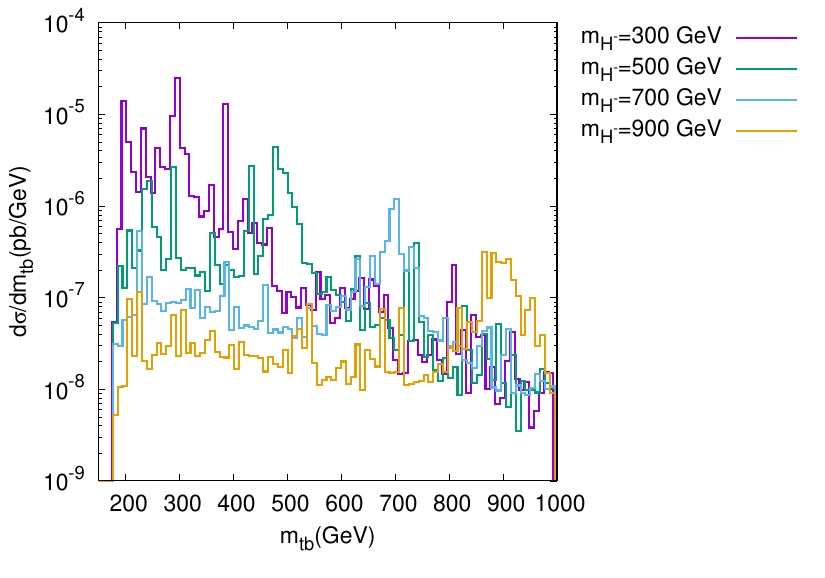}}
\caption{Invariant Mass distributions of tb quarks of THDM-IV.
Fig.a shows $m_{tb}$ data plots of t$\bar{b}$ quarks for BP-I.
Fig.b shows $m_{tb}$ of t$\bar{b}$ quarks for BP-II}
\label{fig:13}
\end{figure}

Figures \ref{fig:12} and \ref{fig:13} display the invariant mass curves of \emph{t}$\bar{b}$ quarks for varying charged Higgs masses in THDM-III and THDM-IV models. A distinct peak in the distribution of $\frac{dN}{dm_o}$ is anticipated at the mass of the undetected particle resulting from the collision. \\
As scaled earlier there will be a distinct peak at the mass of the missing particle. The sharp peaks can be seen at different Higgs masses. We also have plotted the distributions for SM, which shows no sharp peak indicating the charged Higgs presence behind the SM like two Higgs doublet models, for which phenomenology is discussed in
the current studies.
Figure \ref{fig:11} in THDM-IV for BP-II shows significant peaks. The sharper peaks occur at Higgs masses of $M_{H^-}=300$ GeV and $M_{H^-}=900$ GeV, while broader peaks are observed at $M_{H^-}=500$ GeV and $M_{H^-}=700$ GeV, suggesting potential measurement errors and experimental challenges in detection. \\
In Figure \ref{fig:11}(a) for BP-I, the more sharp peak is at $M_{H^-} = 300 GeV$. All the rest are broader peaks as compared to charged Higgs of mass 300 GeV.
\section{Observability of Charged Higgs using Multivariate Analysis}

Multivariate Analysis (MVA) techniques are frequently used in ATLAS and CMS analyses to differentiate signals from backgrounds in searches involving complex multi-particle final states. For instance, in the current analysis of $pp \rightarrow H^{\pm} W^{\mp} \rightarrow t\bar{b} W^{\mp} \rightarrow b\bar{b}W^{\mp}W^{\pm} \rightarrow $ 2 bjets + 2l + MET as the signal channel, electroweak processes are treated as background events that produce nearly identical final states.

Multivariate classification techniques are essential for data analysis, leveraging machine learning algorithms. The TMVA toolkit \cite{tmva}, integrated with the ROOT framework \cite{root}, offers a wide range of classification algorithms. It enables simultaneous training, testing, evaluation, and utilization of various classifiers, making it user-friendly.
TMVA methods use supervised machine learning, where training events are used to determine desired outputs. The algorithm learns to distinguish between background and signal events through a training process analyzing known events, such as simulated samples with predefined outcomes.

%
Three classifiers are used for the charged Higgs analysis in association with the W boson: Boosted Decision Trees (BDT), Maximum Likelihood (LH) method, and Multilayer Perceptron (MLP). The production samples are generated using Monte Carlo (MC) simulations of the signal and various backgrounds at a center-of-mass energy of $\sqrt{s}$ = 100 TeV. The main backgrounds include single top quark production in association with the W boson, single top production in the s-channel, top anti-top quark pair production, and their purely leptonic decays.


In Type I 2HDM, one Higgs doublet provides mass to gauge bosons and fermions, while the other Higgs doublet contributes through mixing.
The Higgs doublet $\phi_2$ couples with all quarks (up-type and down-type) and charged leptons, naturally conserving flavor. This is one of two discrete scenarios that achieve natural flavor conservation in 2HDM. The model under consideration requires the calculation of various parameters in the physical basis, including $m^2_{12}$, tan $\beta$, physical Higgs masses $(m_{H^\pm}, m_{A}, m_{h}, m_{H)}$, mixing angle $\alpha$, sin($\beta - \alpha)$, $\lambda_6$, and $\lambda_7$.\\
In this, we operate close to the alignment limiting, where sin($\beta - \alpha)$ = 0.997806.  
The constraint assumes that the lighter CP-even Higgs boson (h) behaves like the Standard Model Higgs boson with a mass of 125.1 GeV. The heavy neutral Higgs mass is set to 308.86 GeV, the pseudoscalar Higgs boson mass to 411.99 GeV, and the charged Higgs mass to 272.13 GeV. Operating at a center-of-mass energy of 100 TeV with tan$\beta$=3.66 simplifies the calculation and analysis of the model's properties. Signal and background processes are generated using Pythia8 \cite{pythia8} in MadGraph \cite{madgraph}, and Delphes \cite{delphes} simulates the detector. The event generation utilizes the default parton density functions (PDFs) embedded in the Monte Carlo generator. To simulate detector effects, Delphes is run using one of the standard configuration cards supplied with the framework. This setup allows for quick and consistent simulation of particle collision events and their detector responses.\\
We set $m_h$ = 125.1 GeV and randomly vary the other 2HDM parameters within specified ranges:
126 $\le m_H \le $ 500 GeV, 60 $\le m_A \le$ 500 GeV, 80 $\le m_{H^\pm} \le$ 400 GeV, 2 $\le tan\beta \le$ 20, 0.95 $\le sin(\beta - \alpha)$ $\le$ 1 and $m^2_{12} = m^2_H$ sin$\beta$ cos$\beta GeV^2$.
All parameter points must adhere to theoretical and experimental constraints. The 2HMDC \cite{2hdmc} enforces theoretical constraints such as perturbativity, unitarity, and vacuum stability, as well as Electroweak precision constraints using oblique parameters S, T, and U. Flavor constraints are applied using SuperIso v4.1 \cite{superiso}, and exclusion limits at a 95\%  confidence level from BSM Higgs searches at collider experiments are determined using the latest version of HiggsBounds \cite{Higgsbound} through HiggsTools \cite{Higgstools}.

\begin{figure}
\centering     
\includegraphics[scale=0.5]{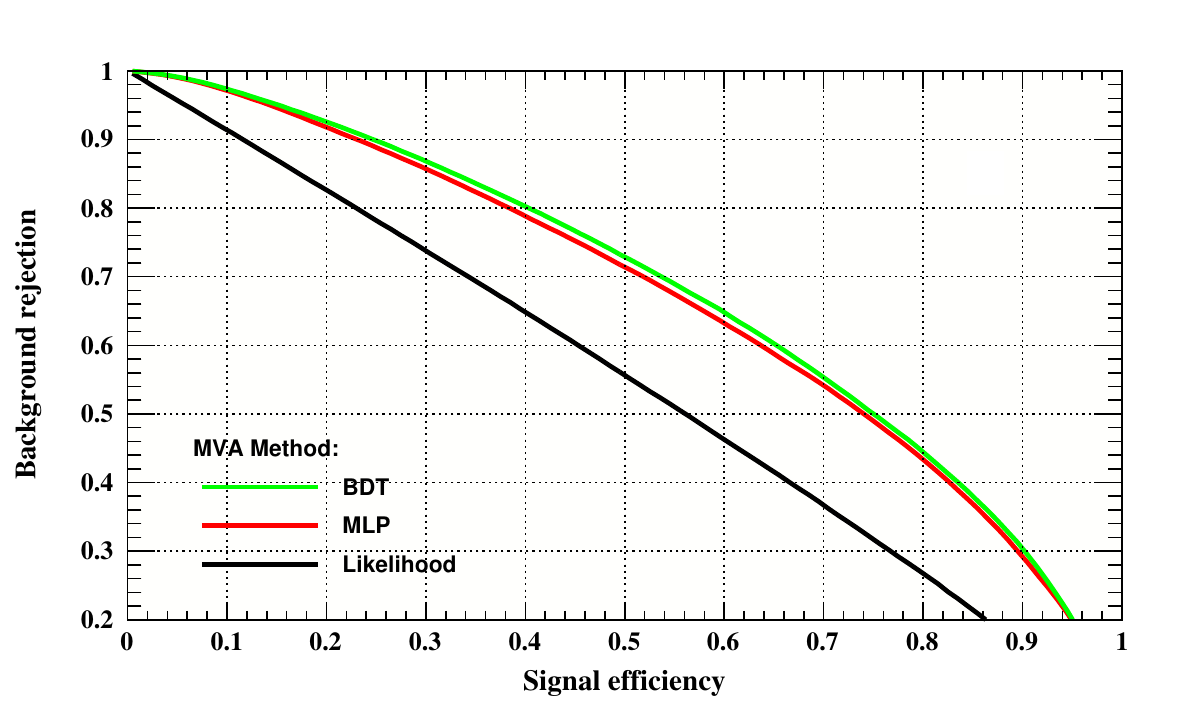}
\caption{A view of the signal efficiency and background rejection curve.}
\label{roc}
\end{figure}

\begin{table}[]
\def\arraystretch{1.7}
    \centering
\begin{tabular}{||c| c| c| c| c||}\hline\hline
 \textbf{MVA  Classifier} & \textbf{Optimal-Cut} & $\mathbf{\dfrac{S}{\sqrt{S+B}}}$ & \textbf{Sig-Eff}&\textbf{Bkg-Eff} \\\hline
       \textbf{MLP}&0.292&160.41&0.972&0.865\\\hline 
           \textbf{Likelihood} &-0.272&158.37&0.990&0.964\\\hline
\textbf{ BDT}&-0.140&160.36&0.967&0.849\\\hline     
    \end{tabular}
   \caption{The Optimal-Cuts, Signal-Background ratio, signal, and Background efficiency for the number of signal and background with applying Cuts.}
    \label{cuts}
\end{table}
The signal and background samples are analyzed using the Toolkit for Multivariate Analysis in ROOT with different classification algorithms. Events are chosen based on the required number of objects in the final state. Signal events are identified by having at least 2 b-jets, 2 leptons, missing transverse energy below 120 GeV, lepton transverse momentum above 20 GeV, and absolute pseudorapidity under 2.1. Selection cuts are used to reduce background and improve signal detection.

The Receiver Operating Characteristic (ROC) curve is a valuable tool for evaluating classifier performance as it visually represents the trade-off between sensitivity and specificity.
By examining the curve, one can determine the optimal threshold for classification, which is the point at which the model is most effective at distinguishing between positive and negative classes. 
Figure \ref{roc} displays the responses of all three classifiers. The Area Under the Curve (AUC) values are as follows: MLP=0.67, BDT=0.70, Likelihood=0.6. These values indicate moderate to satisfactory performance, with room for improvement in the Likelihood method in future iterations.
We calculate statistical significance (SS) using the signal (S) and background (B) events, defined as $SS= S /\sqrt{S+B}$ as shown in Table \ref{cuts}.
The observability of the charged Higgs at the FCC-integrated luminosity depends on the signal efficiency at the ideal cut value. Low signal efficiency can result in filtering out many signals due to statistical limitations. In cases of limited statistical data, the cut quality may be compromised as the output distribution may differ from the training sample. Figure \ref{fig:mlp} shows the MLP cut efficiencies based on the cut value applied. MLP and BDT demonstrate higher significance compared to the Likelihood method, with both methods producing similar results.

\begin{figure}
\centering     
\subfigure[]{\label{fig:a}\includegraphics[scale=0.4]{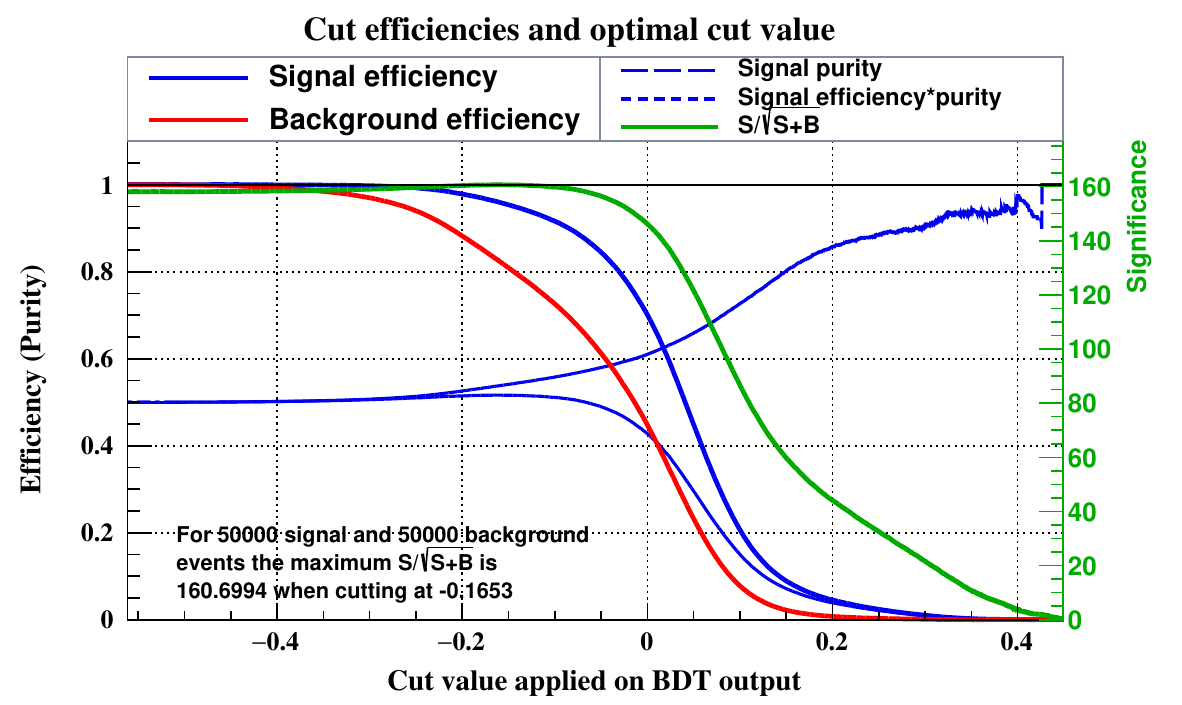}}
\subfigure[]{\label{fig:a}\includegraphics[scale=0.4]{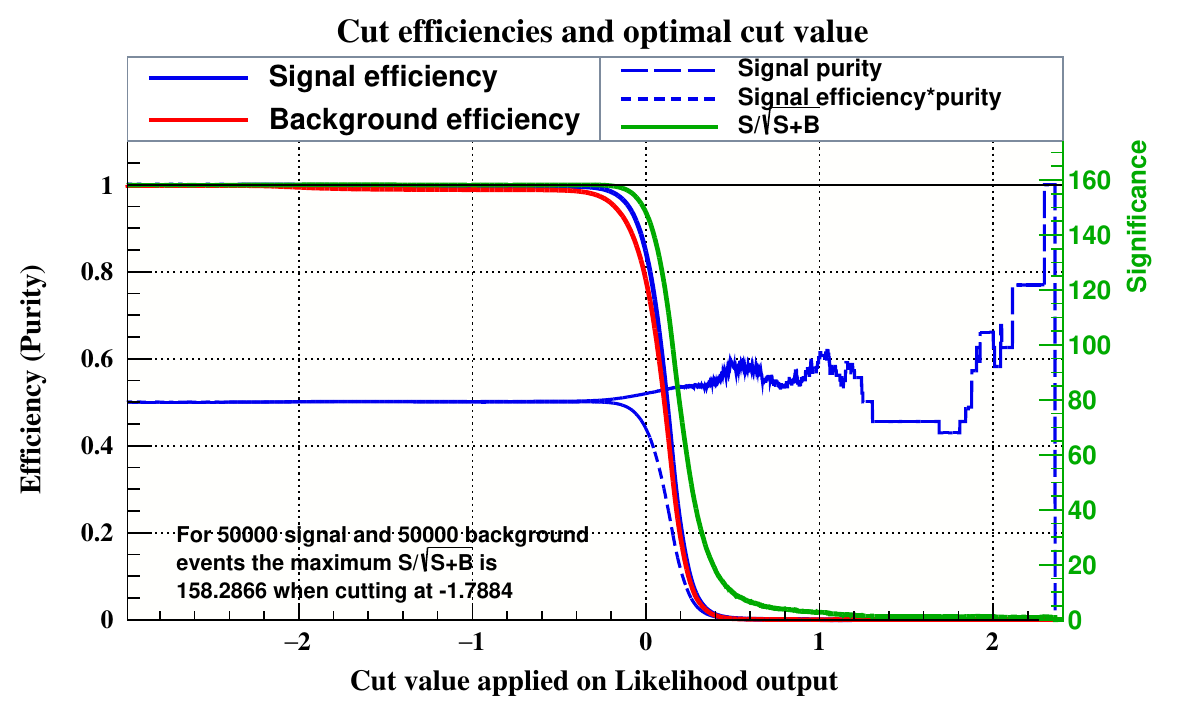}}
\caption{Signal efficiency, background efficiency, signal purity, signal efficiency multiplied by purity, and significance $S/\sqrt{S+B}$ are plotted against the BDT cut level in (a) and Likelihood output level in (b)
as a function of the BDT cut value.}
\label{fig:bdtlh}
\end{figure}

\begin{figure}
\centering     
\includegraphics[scale=0.5]{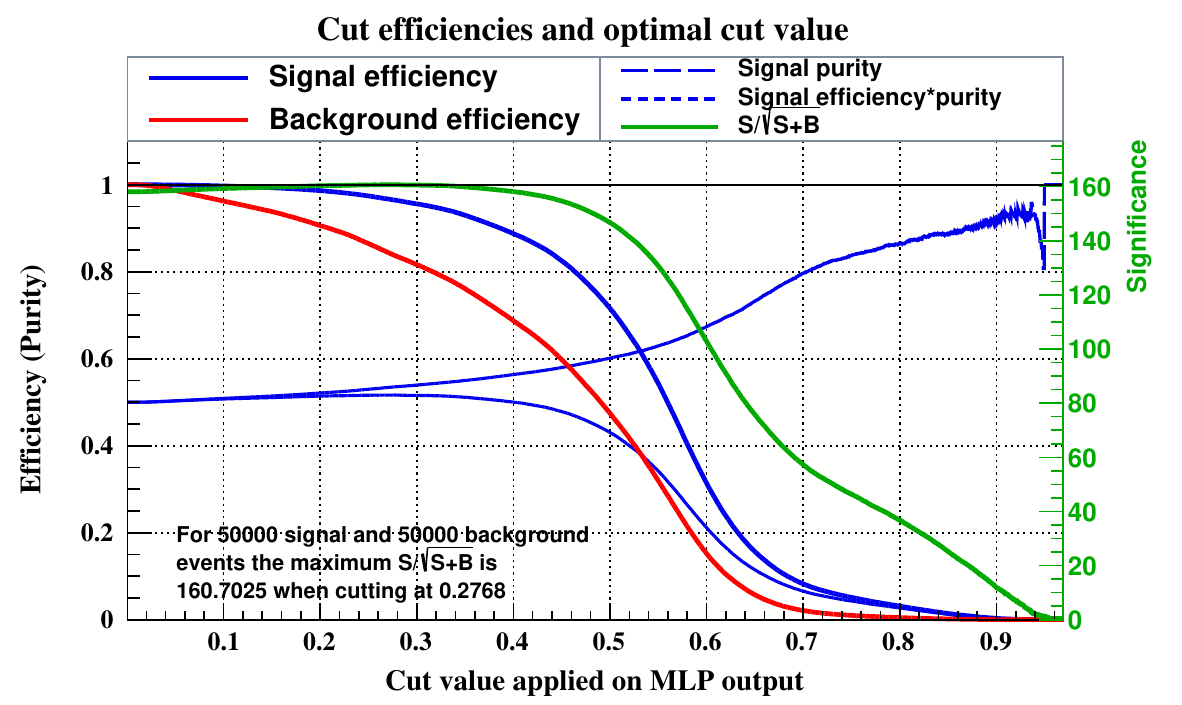}
\caption{The MLP cut efficiencies as a function of applied cut values.}
\label{fig:mlp}
\end{figure}

Figure \ref{fig:bdtlh} shows the significance $S/\sqrt{S + B}$ as a function of the BDT cut value in (a) and the Likelihood cut in (b) for a fixed number of signal and background events. This metric accurately assesses the classifier's performance, with higher significance indicating better performance.
However, it is crucial to consider the signal efficiency corresponding to the optimal cut value. If the signal efficiency is too low, a significant amount of signal may be lost.

The quality of the cut may vary if the actual data has poor statistics, leading to differences in the output distribution from the training sample. Figure \ref{fig:mlp} illustrates the MLP cut efficiency based on the applied cut value. Both MLP and BDT methods demonstrate comparable efficiency, making them more pertinent than the Likelihood approach.

\section{Conclusion}
A comprehensive analysis has been conducted to predict the observability of charged Higgs across a broad range of parameter space in the extended Higgs sector in the context of extremely high energy proton-proton collisions. The production and decay properties of charged Higgs rely on the ongoing searches of LHC. We have concentrated on the 2HDM model type III \& IV scenario, studied the charged Higgs boson, and top quark associated production in proton-proton collisions at the FCC energy. \\
Multivariate Analysis methods are commonly utilized in ATLAS and CMS to distinguish signals from backgrounds in searches involving complex multi-particle final states. The TMVA toolkit, integrated with the ROOT framework, provides a range of classification algorithms.

%
Three classifiers (BDT, LH, and MLP) were selected to analyze the charged Higgs in association with a W boson. Production samples were generated through Monte Carlo simulations of signal and background events at a center-of-mass energy of 100 TeV.
The study focuses on single top quark production, single top production in the s-channel, and top anti-top quark pair production with purely leptonic decays. The model under investigation is a Type I 2HDM, where mass is acquired by gauge bosons and fermions from one Higgs doublet, while the other Higgs doublet contributes to mass through mixing. Model parameters were scanned within specific ranges, and signal and background samples were analyzed using the TMVA toolkit.

%
The Receiver Operating Characteristic (ROC) curve was used to evaluate the performance of the classifiers, and the results showed that the MLP and BDT methods performed better than the LH method. 
The statistical significance (SS) was calculated for signal and background events. It was observed that MLP and BDT had higher significance than LH.
%



\begin{thebibliography}{9}
\bibitem{z1} Aad, G., et al, Observation of a new particle in the search for the Standard Model Higgs boson with the ATLAS detector at the LHC, Phys. Lett. B716, C29 (2012). arXiv:1207.7214[hep-ex]

\bibitem{z2} Chatrchyan, S., et al, Observation of a new boson at a mass of 125 GeV with the CMS experiment at the LHC, Phys. Lett. B716, 361
(2012). arXiv:1207.7235[hep-ex]

\bibitem {z3} Steven D. Bass, Albert De Roeck, and Marumi Kado, The Higgs boson implications and prospects for future
discoveries, Nature 3, 608 (2021).

\bibitem{z4} A. G. Akeroyd et al, Prospects for charged Higgs searches at the LHC, Eur. Phys. J. C (2017) 77:276

\bibitem {z5} G. C. Branc et al, Theory and Phenomenology of two-Higgs-doublet models, Physics Reports 516 (2012) 1102
    
\bibitem {z6} I. T. Cakir et al, Probing charged Higgs boson couplings at a future circular hadron collider, Phys. Rev. D 94, 015024 (2016)

\bibitem{z7} Shuailong Li, Huayang Song and Shufang Su, Probing exotic charged Higgs decays in the Type-II 2HDM through top rich signal at a future 100 TeV pp collider, JHEP
11, 105 (2020).

\bibitem{z8} Vernon Barger, Heather E. Logan, and Gabe Shaughnessy, Identifying extended Higgs models at the LHC, Phys. Rev. D 79, 115018 (2009)

\bibitem {z9} Ijaz Ahmed, Sehrish Gul, Taimoor Khurshid Received, Correlation Between the Scaling Factor of the Yukawa Coupling and Cross Section for the $e^+e^-\rightarrow hhff_{bar} (f\neq t ) in Type-I 2HDM$, Int J Theor Phys 60, 2916-2929 (2021).

\bibitem{z10} A. Arhrib et al, Extended Higgs sector of 2HDM with real singlet facing LHC
data, Eur. Phys. J. C (2020) 80:13.

\bibitem{z11} Rodolfo A. Diaz, R. Martinez, and J.-Alexis Rodriguez, Lepton flavor violation in the two Higgs doublet model
type III, Phys. Rev. D, VOLUME 63, 095007 (2001).

\bibitem{z12} A. Arhrib et al, Two-Higgs-doublet type-II and -III models and $t\rightarrow ch$ at
the LHC, Eur. Phys. J. C (2016) 76:328.

\bibitem {z13} R. Primulando, J. Julio and P. Uttayarat, Collider constraints on lepton flavor violation in the
2HDM, Phys. Rev. D 101, 055021 (2020)

\bibitem{52}
    ALEPH, DELPHI, L3, OPAL, LEP collaboration, G. Abbiendi et al., Search for Charged Higgs bosons: Combined Results Using LEP Data, Eur. Phys. J. C73, 2463 (2013). arXiv:1301.6065
    
\bibitem{fccweb} The Future Circular Collider". 18 June 2025. https://home.cern/science/accelerators/future-circular-collider

\bibitem{abada} A. Abada et al., FCC Physics Opportunities, Future Circular Collider Conceptual Design Report Volume 1, The European Physical Journal C, Volume 79, article number 474, (2019)

\bibitem{fccbeam} https://cds.cern.ch/record/2651300/files/CERN-ACC-2018-0058.pdf pg. 248, Beam Parameters gives GJ of total energy based on number of protons per bunch and number of bunches [10,400] in FCC-hh


    
\bibitem{54}
    ATLAS collaboration, G. Aad et al., Search for charged Higgs bosons decaying vi $H^{\pm}\rightarrow\tau^{\pm}\nu$ fully hadronic final states using pp collision data at $\sqrt{s}$=8 TeV with the ATLAS detector, JHEP 03, 088 (2015). arXiv:1412.6663
\bibitem{55}
    CMS collaboration, V. Khachatryan et al., Search for a charged Higgs boson in pp collisions at  $\sqrt{s}$=8 TeV JHEP 11, 018 (2015). arXiv:1508.07774
\bibitem{56}
    CMS collaboration, V. Khachatryan et al., Search for a charged Higgs boson in pp collisions at  $\sqrt{s}$=8 TeV JHEP 11, 018 (2015). arXiv:1508.07774
\bibitem{57}
    ATLAS collaboration, G. Aad et al., Search for charged Higgs bosons decaying vi $H^{\pm}\rightarrow\tau^{\pm}\nu$ fully hadronic final states using pp collision data at $\sqrt{s}$=8 TeV with the ATLAS detector, JHEP 03, 088 (2015). arXiv:1412.6663
 
 \bibitem{calchep1}
 Pukhov, A., et al. "CalcHEP: A package for calculation of Feynman diagrams and integration over multi-particle phase space." hep-ph/9908288. 

 \bibitem{calchep2}
 Belyaev, A., et al. "CalcHEP 3.4 for collider physics within and beyond the Standard Model." arXiv:1207.6082
 
 \bibitem{gnuplot}   
    Williams, T. and Kelley, C. (2011). Gnuplot 4.5: an interactive plotting program. URL http://gnuplot.info.
    
\bibitem{58}
    Prasenjit Sanyal,' Limits on the charged Higgs parameters in the two Higgs doublet model using CMS $\sqrt{s}$=13 TeV results',  The European Physical Journal C volume 79, Article number: 913 (2019).
\bibitem{51}
    M. Guchait and A. H. Vijay, “Probing Heavy Charged Higgs Boson at the LHC,” Phys. Rev., Vol. D98, No. 11, p. 115028, 2018.
\bibitem{67}
    David J. Griffiths, Introduction to elementary particles. John Wiley \& Sons.Inc,1987.
\bibitem{66}
    Cheuk-Yin Wong, “Introduction to High-Energy Heavy-Ion Collisions", World Scientific Publishing Co. Pte. Ltd

\bibitem{tmva} A. Hocker, P. Speckmayer, J. Stelzer, J. Therhaag, E. von Toerne, H. Voss, M. Backes, T. Carli, O. Cohen and A. Christov, et al. arXiv:physics/0703039.

\bibitem{pythia8} T Sjöstrand, Torbjorn, Stefan Ask, Jesper R. Christiansen, Richard Corke, Nishita Desai, Philip Ilten, Stephen Mrenna, Stefan Prestel, Christine O. Rasmussen, and Peter Z. Skands. ”An introduction to PYTHIA 8.2.” Computer Physics Communications 191 (2015): 159-177.
\bibitem{madgraph}J. Alwall, M. Herquet, F. Maltoni, O. Mattelaer, and T. Stelzer, ``MadGraph 5: going beyond", JHEP 06, 128 (2011), arXiv:1106.0522, (2011).
\bibitem{delphes}. de Favereau, C. Delaere, P. Demin, A. Giammanco, V. Lemaître, A. Mertens, and M. Selvaggi. DELPHES 3, A modular framework for fast simulation of a generic collider experiment. JHEP, 02:057, 2014
\bibitem{2hdmc}  D. Eriksson, J. Rathsman and O.Stal, 2HDMC - Two-Higgs-Doublet Model Calculator., Comput. Phys. Commun. 181 (2010) 189.     
\bibitem{superiso} F. Mahmoudi, SuperIso v2.3: A program for calculating flavor physics observables in supersymmetry, Computer Physics Communications, Volume 180, Issue 9, 2009, Pages 1579-1613,
\bibitem{Higgsbound} P. Bechtle et al., Comput.Phys.Commun.181:138-167,2010     
\bibitem{Higgstools} P. Bechtle et al., Comput. Phys. Commun. 182 (2011), 2605-2631
\bibitem{root} Rene Brun and Fons Rademakers, ROOT - An Object Oriented Data Analysis Framework, Proceedings AIHENP 96 Workshop, Lausanne, Sep. 1996, Nucl. Inst. Meth. in Phys. Res. A 389 (1997) 81-86.
\end{thebibliography}
\end{document}